\documentclass[11pt,twoside]{article}
\usepackage{asp2010}

\resetcounters

\begin{document}

\title{The mysterious sdO X-ray binary BD+37$^\circ$442}
\author{U. Heber$^1$, S. Geier$^{1,2}$,  A. Irrgang$^1$,  D. Schneider$^1$, I. Barbu-Barna$^1$, S. Mereghetti$^3$, N. La Palombara$^3$
\affil{$^1$Dr. Karl Remeis-Observatory \& ECAP, Astronomical Institute,
Friedrich-Alexander University Erlangen-Nuremberg, Sternwartstr. 7, D 96049 Bamberg, Germany}
\affil{$^2$European Southern Observatory, Karl-Schwarzschild-Str. 2, D-85748 Garching, Germany}
\affil{$^3$INAF, Istituto di Astrofisica Spaziale e Fisica Cosmica Milano, via E. Bassini 15, 20133 Milano, Italy}}

\begin{abstract} Pulsed X--ray emission in the luminous, helium-rich sdO BD\,+37$^\circ$442 has recently been discovered (La Palombara et al., 2012). It was suggested that  the sdO star has a neutron star or
white dwarf companion with a spin period of 19.2 s. After HD 49798, which has a massive
white dwarf companion spinning at 13.2 s in an 1.55 day orbit, this is only the
second O-type subdwarf from which X-ray emission has been detected.
We report preliminary results of our ongoing campaign to obtain time-resolved high-resolution spectroscopy using the CAFE instrument at Calar Alto observatory and SARG at the Telescopio Nationale Galileo. Atmospheric parameters were derived via a quantitative NLTE spectral analysis. The line fits hint at an unusually large projected rotation velocity.
Therefore it seemed likely that BD\,+37$^\circ$442 is a binary similar to HD 49798 and that the orbital period is also similar. The level of X-ray emission from BD\,+37$^\circ$442 could be explained by accretion from the sdO wind by a neutron star orbiting at a period of less than ten days.
Hence, we embarked on radial velocity monitoring in order to derive the binary parameters of the BD$+$37$^\circ$442 system and obtained 41 spectra spread out over several month in 2012. Unlike for HD\,49798, no radial velocity variations were found and, hence, there is no dynamical evidence for the existence of a compact companion yet. The origin of the pulsed X-ray emission remains as a mystery.  

\end{abstract}

\section{X-rays from sdO stars}
\begin{figure*}
\begin{center}
 \includegraphics[width=4cm,angle=270]{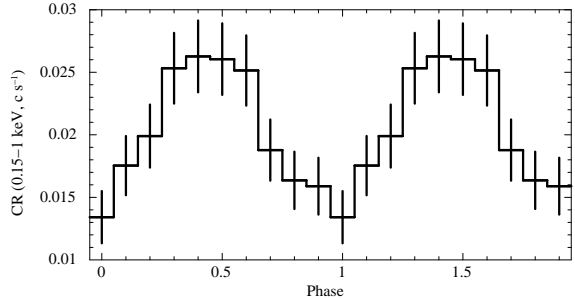}
 \caption{X-ray  light curve of BD+37$^\circ$442 in the 0.15-2 keV energy range,
folded at the period P=19.156 s (La Palombara et al., 2012)}
 \label{lc}
\end{center}
\end{figure*}

The compact companions of hot subluminous stars
are difficult to detect in optical/UV data. However, they might give rise to 
detectable X-ray emission if the subdwarf
can provide an adequate accretion rate. 
In this respect, luminous sdO stars offer the best prospects for X-ray detections because of their high mass loss rates ($\dot M \sim10^{-7}-10^{-10} M_{\odot}$ yr$^{-1}$; Jeffery \& Hamann, 2010).
This is well demonstrated by the case of the luminous sdO HD 49798, a single-lined
spectroscopic binary (P=1.5 days) from which soft X--rays  were 
discovered  with the ROSAT satellite. Remarkably, the X-ray flux shows a strong periodic
modulation at 13.2 s  (Israel et al., 1997), indicating the presence of a compact object. 
X-ray eclipses (which constrains the system inclination) allowed a dynamical measurement of the masses of the two binary components.
This showed that the companion of HD 49798 is a very
massive (1.28$\pm$0.05 $M_{\odot}$) white dwarf (Mereghetti et al. 2009, 2011a, 2013),
and qualifies this binary as a potential progenitor of a type Ia supernova.
Prompted by the case of HD 49798, a search for X-ray emission from other hot subdwarfs was launched.
While searches with the Swift/XRT of candidate sdB+WD binaries gave negative results (Mereghetti et al., 2011b),
pulsed X--ray emission in the luminous, hydrogen-deficient sdO BD\,+37$^\circ$442 was discovered (La Palombara et al., 2012). The observed X-rays have a soft spectrum (a blackbody with temperature $\sim$45 eV
plus a weak power-law component), and
show a significant modulation at a period of 19.2 s (see Fig \ref{lc}).

The best fit X-ray luminosity is $\sim10^{33}$ erg s$^{-1}$ (for a distance of 2 kpc; Bauer \& Husfeld, 1995),
but, due to the uncertainties in the X-ray blackbody temperature, the bolometric X-ray luminosity
could be between $\sim10^{32}$  and $\sim10^{35}$ erg s$^{-1}$ (see Fig.2).
The detection of pulsations indicates the presence of a WD or a neutron star, 
most likely powered by accretion from the stellar wind of BD+37$^\circ$442,
which is losing mass at a rate of $\sim$3$\times$10$^{-9}$ $M_{\odot}$   yr$^{-1}$ and with
a wind terminal velocity of v$_{\infty}$=2,000\,km\,s$^{-1}$ (Jeffery \& Hamann, 2010). 
Assuming a canonical wind velocity law and Bondi-Hoyle accretion, La Palombara et al. (2012)
showed that a neutron star orbiting BD+37$^\circ$442 with a period of less than ten days (for a circular orbit) would accrete from the wind
at a sufficiently high rate to produce the observed X-rays. The companion could be a white dwarf only if the system were much tighter allowing Roche-lobe overflow.

\section{Optical spectroscopy}

We obtained 41 high-resolution spectra in the second half of 2012 using the CAFE spectrograph at Calar Alto observatory (27 spectra) and the SARG spectrograpgh at the TNG (La Palma, 14 spectra).

\subsection{Atmospheric parameters and projected rotation velocity of BD+37$^\circ$442}

The optical spectrum of BD+37$^\circ$442 is dominated by \ion{He}{II} lines, with no traces of hydrogen Balmer line blends. Many lines of carbon (\ion{C}{iii} and \ion{C}{IV}) can be identified hinting at a carbon-rich composition. A few \ion{C}{iii} lines are in emission, most notably $\lambda$ 5695.92\AA. We adopted a He/H ratio of 100:1 by number and carried out a quantitative spectral analysis of the helium line spectrum using NLTE model atmospheres, which resulted in an effective temperature T$_{\rm eff}$ = 56300\,K 
and surface gravity $\log g$ = 4.1. Additional line broadening has to be invoked in order to match the observed line profiles by synthetic ones which hinted at a projected rotational velocity of 60 km\,s$^{-1}$. However, because we used static models, the line broadening could be (partly) due to dynamic phenomena such as macroturbulence or the stellar wind.  

Most other hot subdwarfs are known to be slow rotators with projected rotational velocities of less than 30\,km\,s$^{-1}$ ( Hirsch \& Heber, 2009) for apparently single sdO stars and less than 10\,km\,s$^{-1}$ (Geier \& Heber, 2012) for single sdB stars. Hence  BD+37$^\circ$442 may rotate faster than expected. However, the sdO star could have been spun up by a compact companion. If the rotation of the sdO star should be tidally locked to its orbit, the period must be of the order of 1 day.

\subsection{A search for radial velocity variations}

\begin{figure*}
\begin{center}
 \includegraphics[width=9cm]{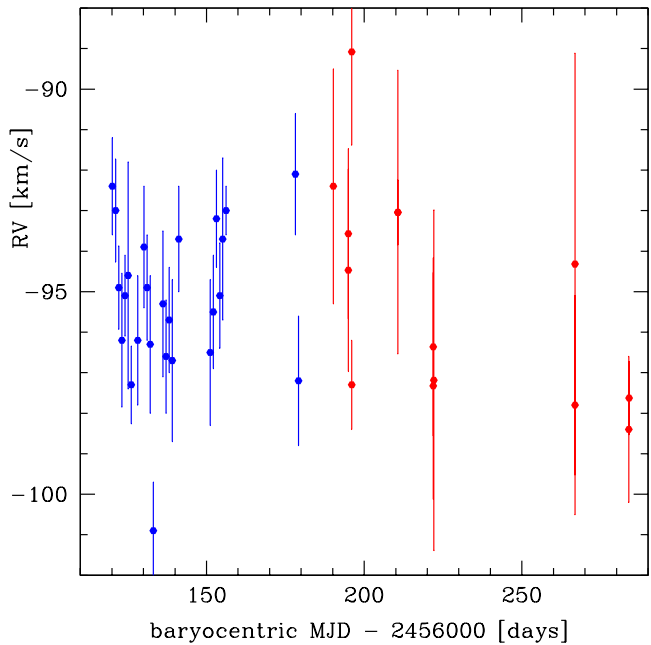}
 \caption{Radial velocities  of BD+37$^\circ$442 derived from CAFE (before MJD 2456185, blue) and TNG/SARG spectra (after MJD=2456185, red).}
 \label{rv}
\end{center}
\end{figure*}

The orbital period of BD+37$^\circ$442 is unknown since radial velocity measurements are scarce. Drilling \& Heber (1987) report a heliocentric radial velocity of $-$94$\pm$1\, km\,s$^{-1}$. Radial velocity variations have not been noted in the literature. As outlined above, we expect the orbital period to be of the order of a few hours to ten days. Therefore we embarked on a time series of optical spectra.  Radial velocities were derived from helium, carbon and nitrogen lines and are displayed in Fig.~\ref{rv}. No radial velocity variations were found on the time scales of hours, days, weeks or months at a level of few km\,s$^{-1}$. The mean velocity is perfectly consistent with 
that measured by Drilling \& Heber (1987).

\section{Discussion}

The level of X-ray emission from BD\,+37$^\circ$442 could be explained by accretion from the sdO wind by a neutron star orbiting at a period of less than ten days. Therefore it seemed likely that BD\,+37$^\circ$442 is a binary similar to HD 49798 and that the orbital period is also similar to that of the latter (1.55 d). If the companion to BD\,+37$^\circ$442 were a white dwarf, short period variations are expected to explain the level of the X-ray flux through accretion via Roche lobe overflow. Hence the companion is more likely a neutron star.
However, our time series spectroscopy did not reveal any radial velocity variations at the expected period and amplitude range. 
Possible explanations for the lack of RV variations are   
\begin{itemize}
\item[1.)]  BD$+$37$^\circ$442 is not a binary at all. Then the origin of the X-ray pulses remains a mystery. The similarity to HD~49798, however, suggests that BD$+$37$^\circ$442 is a binary too, and the pulsed X-ray emission is due to rotation of the accreting neutron star.
\item[2.)] The orbital plane of BD$+$37$^\circ$442 is almost perpendicular to the line of sight. In this case the rotation axes of the sdO star as well as of the X-ray emitting companion must be strongly inclined w.r.t. the orbital plane.
\item[3.)] The orbital period is very long (years) and the orbit strongly eccentric. A radial velocity campaign of much longer duration would be needed. X-ray emission is predicted to occur close to periastron only.  
\end{itemize}
This calls for concerted, long-term spectroscopic monitoring and repeated X-ray observations.

\acknowledgements Based on observations collected at the Centro Astronómico Hispano Alemán (CAHA) at Calar Alto, operated jointly by the Max-Planck Institut für Astronomie and the Instituto de Astrofísica de Andalucía (CSIC)' and
on observations made with the Italian Telescopio Nazionale Galileo (TNG) operated on the island of La Palma by the Fundaci\'on Galileo Galilei of the INAF (Istituto Nazionale di Astrofisica) at the Spanish Observatorio del Roque de los Muchachos of the Instituto de Astrofisica de Canarias". AI is supproted by Deutsche Forschungsgemeinschaft through grant He1356/45-2.

\end{document}